\documentclass[preprintnumbers,amsmath,amssymb,twocolumn]{revtex4}
\usepackage[dvips,colorlinks=true,citecolor=red,filecolor=green,linkcolor=blue,pdfnewwindow=true]{hyperref}
\usepackage{graphicx}
\usepackage[title]{appendix}

\begin{document}
\title{Fluctuations in network dynamics: $SMAR1$ can trigger apoptosis}
\author{Md. Zubbair Malik$^{1,4}$, Md. Jahoor Alam$^2$, Romana Ishrat$^1$, Subhash M. Agarwal$^{3}$ and R.K. Brojen Singh$^{4}$}
\email{brojen@jnu.ac.in (Corresponding author)}
\affiliation{$^1$Centre for Interdisciplinary Research in Basic Sciences, Jamia Millia Islamia, New Delhi-110025, India.\\
$^2$College of Applied Medical Sciences, University of Ha’il, Ha’il-2440, Kingdom of Saudi Arabia.\\
$^3$Bioinformatics Division, Institute of Cytology and Preventive Oncology, 1-7, Sector - 39, Noida 201301, India.\\
$^4$School of Computational and Integrative Sciences, Jawaharlal Nehru University, New Delhi-110067, India.}

\begin{abstract}
$SMAR1$ is a sensitive signaling molecule in $p53$ regulatory network which can drive $p53$ network dynamics to three distinct states, namely, stabilized (two), damped and sustain oscillation states. In the interaction of $p53$ network with $SMAR1$, $p53$ network sees $SMAR1$ as a sub-network with its new complexes formed by $SMAR1$, where $SMAR1$ is the central node, and fluctuations in $SMAR1$ concentration is propagated as a stress signal throughout the network. Excess stress induced by $SMAR1$ can drive $p53$ network dynamics to amplitude death scenario which corresponds to apoptotic state. The permutation entropy calculated for normal state is minimum indicating self-organized behavior, whereas for apoptotic state, the value is maximum showing breakdown of self-organization. We also show that the regulation of $SMAR1$ togather with other signaling molecules p300 and HDAC1 in the $p53$ regulatory network can be engineered to extend the range of stress such that the system can be save from apoptosis.
\end{abstract}


\maketitle

\section{Introduction}

More than three decades of its discovery, $p53$ protein is still an important and critical molecule to study to explore new insights of cellular functional organization. Several experimental work have been done on $p53$ to understand how it regulates various cellular functions, but system level organization of these functional pathways controlled by $p53$ in normal, stress and cancerous cells are still to be investigated rigorously to understand the role of $p53$ at various cellular states at fundamental level. Due to its importance in cellular mechanisms, it is nominated as molecule of the year and also molecule of the month by the science magazine in 1993. $p53$ is composed of 393 amino acids \cite{bai}. It has a very short half-life of 15-30 minutes \cite{fin}. It takes part in many important cellular processes such as cell differentiation, maintaining genome integrity, apoptosis \cite{lev} etc. One of the most important negative regulator of the $p53$ is $Mdm2$ protein \cite{lev,kub}. $Mdm2$ forms the complex with $p53$ and then degraded the complex through its enzymatic activity. $p53$ acts as transcription factor for various important signaling molecules which participate in several important cellular networks and pathways. It helps in the formation of $Mdm2$ protein through positive feedback, but in turn $Mdm2$ negatively regulates $p53$ \cite{gro} to maintain minimum $p53$ level at normal state. These feedbacks lead to the oscillatory behavior of $p53$ in the regulatory network system. Further, $p53$ is very sensitive molecule which is generally activated due to several types of cellular stresses, namely DNA damage, interacting with various signaling molecules such as nitric oxide (NO), reactive oxygen synthase (ROS), calcium etc. Once cell comes under the influence of stress the inactive form of ATM kinase get activated and this activated ATM sent the damage information to $p53$ by interacting with it \cite{pro}. Consequently this leads to the phosphorylated $p53$, and encounters to several statges of reactions network to repair DNA damage and comes back to normal state, otherwise move to apoptosis \cite{lam,wag}. 
   
p300 is an acetylating agent which acetylate $p53$ at its c-terminal and this leads to the prevention of $p53-Mdm2$ interaction and this activity leads to the suppression of $p53$ degradation \cite{gro1,gu}. It is also reported that p300 also interact with $Mdm2$ protein to form $p300-Mdm2$ complex as a result of which the level of $Mdm2$ is decreased in the system \cite{kob,ito,li}.

On the other hand, HDAC1 is a deacetylating agent which deacetylate the acetylated form of $p53$. The deactylation of $p53$ by HDAC1 is indirect which occurs due to the recruitment of $Mdm2$ by HDAC1 \cite{ito1}. However, the deacetylated form of $p53$ is very vunrable and comes easily in contact with $Mdm2$, which leads to the degradation of $p53$ \cite{ito1,jua}.

$SMAR1$, a $p53$ target gene, is very a versatile molecule as reported by several experimental study so far \cite{pav,sin,sing}. It can interact with $p53$, $Mdm2$ as well p300 molecule with different affinity \cite{sin}. It is reported that it enhances $p53$ transcriptional activity and stability of $p53$ \cite{pav,sin}. Moreover, it also shows negative impact on both $Mdm2$ and $p300$ \cite{sin}. This molecule is expressed upon DNA damages in a $p53$ dependent manner \cite{sing}. It is also indicates that the interaction of $SMAR1$ with $p53$ in the nucleus helps in stabilizing the $p53$ by displacing its negative regulator $Mdm2$ \cite{jal}. Further, $p53$ has been shown to be deacetylated by its interaction with $Mdm2$ through recruitment of HDAC1 \cite{ito1}. It can also interact with and deacetylate $p53$ by recruiting HDAC1 \cite{yan}. However, knockdown of HDAC1 only partially rescues $p53$ acetylation suggesting that $SMAR1$ employs supplemental mechanisms to regulate $p53$ acetylation \cite{sinh}. Hence, $SMAR1$ can be considered as an important nuclear matrix binding transcription factor which acts as a repressor by recruiting HDAC1 \cite{ram}. 

There have been several mathematical models constructed from the $p53$ regulatory network by taking care of various feedback mechanisms in the network \cite{baro,wag,hunz}, incorporating radiotherapy \cite{shao}, by taking into account apoptosis inhibitors (caps3, caps9) showing various states such as bistability \cite{lege}, considering DNA damage via irradiation \cite{pro,cili}, modeling apoptosis from stress $p53$ \cite{bose,chon}, taking into account pharmacodynamics target such nutilin \cite{pus,Md}, incorporating sometogenesis with Wnt, Axin and nutilin \cite{Md}, embodying signaling molecules such as calcium with NO \cite{alam} and p300 and HDAC1 \cite{aro}, integrating with cell cycle pathway \cite{ala}. These studies show that the introduction of stress in the $p53$ regulatory network allows to switch the stabilized $p53$ state to oscillatory dynamics via DNA damage \cite{cili,alam} and excess stress may lead to apoptosis. It has been observed with evidences that the switching mechanism at different dynamical states of $p53$ correspond to various cellular states. However, the monitoring of specific reactions in the $p53$ regulating network could save the system from apoptosis is an open question. The complexity measurement of various possible states of the system could tell many information inherited in the time series and need to be investigated systematically. The study of role of $SMAR1$ in $p53$ regulation may open up new understanding in the regulatory network, stress management in the stress system, monitoring apoptosis and switching in cancer phase. We present $p53$ regulation driven by $SMAR1$ incorporating p300 and HDAC1 in section II with quasi-steady state approaximation technique and permutation entropy description. Results of simulation of the constructed model with discussion in the section III and conclusion based on the simulation results are described in section IV.

\section{Materials and Methods}

\subsection{$p53-Mdm2-SMAR1$ regulatory model}

$p53$ maintained at low levels in unstressed cell due to $p53$ and $Mdm2$ protein feedback mechanism \cite{kub}. It binds to the $Mdm2$ gene in nucleus which leads to the transcription of $Mdm2$ messenger RNA (mRNA) with a rate constant $k_3$, subsequently this leads to translation into $Mdm2$ protein with a rate constant $k_2$ \cite{pro}. The half life of the $Mdm2-mRNA,Mdm2$ is less which occurs with rate $k_4$, $k_5$ respectively. The synthesis of $p53$ protein in cells varies according to the half life of $p53$ protein. We considered the rate of $p53$ synthesis takes place at the rate $k_6$. The interaction of $p53$ and $Mdm2$ is reported with rate $k_8$ which leads to formation of $Mdm2-p53$ complex. It is reported by several research that $Mdm2$ functions as an E3 ubiquitin ligase and this leads to the degradation of $p53$ protein with rate $k_7$. Further the dissociation of the $Mdm2-p53$ complex occurs with a rate $k_9$. When cell experience stress the inactivated form of ATM transform into activated form $Mdm2$ which supposed to be occurs with a rate $k_{10}$. Further it is reported that activated form tansform into inactivated ATM with a rate $k_{11}$. The activated form of ATM interact with $p53$ which leads to phosphorylation of $p53$ with a rate $k_{12}$. This phosphorylated form of $p53$ further dephosphorylates with a rate $k_{13}$. p300 is an important protein which interact with $p53$ and forms $p53$-p300 complex with a rate $k_{15}$ and this subsequently leads to the production of acytylated $p53$ with a rate $k_{16}$ The synthesis of p300 is reported as rate of $k_{23}$. Simmilarly due to its short half life the degradation of p300 is reported to occur with a rate $k_{14}$. p300 is also interact with a rate $k_{20}$ to form $Mdm2-p300$ complex. Further it is reported that $Mdm2-p300$ complex interact with $p53$ and leads to degradation of complex at rate $k_1$ \cite{gro,gro1}. It also reported that $Mdm2-p53$ complex can interact with p300 to form $Mdm2-$p53$-p300$ ternary complex with rate $k_{19}$. Further the dissociation of this complex leads to the formation $Mdm2$ and $p53$-p300 complex with a rate $k_{21}$. HDAC1 deacetylate acetylated $p53$ with rate $k_{33}$ by recruiting $Mdm2$ with rate $k_{32}$. The synthesis and degradation of the HDAC1 due to its half life occurs with rate $k_{24}$ and $k_{22}$. $SMAR1$ is an important signaling molecule which interact with $p53$ and phosphorylates it with a rate $k_{29}$. $SMAR1$ interact with $Mdm2$ to form $Mdm2-SMAR1$ complex with a rate $k_{18}$. $Mdm2-SMAR1$ complex interact with HDAC1 to form $Mdm2-SMAR1-HDAC1$ complex with rate $k_{25}$. Now this bigger complex interact with acetylated $p53$ which leads to the formation of deacetylated $p53$ with a rate $k_{17}$. The synthesis and degradation of the $SMAR1$ due to its half life occurs with rate $k_{26}$ and $k_{27}$. The degradation of $Mdm2-SMAR1$ complex takes place with rate $k_{28}$. $SMAR1$ interact with p300 and degraded its level with rate $k_{35}$. The interaction of $SMAR1$ with $Mdm2-p53$ complex occurs with a rate $k_{30}$ which leads to the formation of $p53-Mdm2-SMAR1$ complex. Further it is reported that the dissociation of this complex with rate $k_{31}$ leads to the formation of phoshporylated $p53$ and $Mdm2-SMAR1$ complex. It is also reported that $SMAR1$ can interact with $p53-p300$ complex with rate $k_{34}$.  

The stress $p53-Mdm2-SMAR1$ model network we study is defined by $N=18$ (18 molecular species) and $M=35$ (35 reaction channels). The molecular species, possible reactions, kinetic laws and the rate constants in this model are listed in Table 1 and Table 2 respectively. The state vector at any instant of time $t$ is given by, ${\vec x}(t)=(x_1,\dots,x_{18})^T$, where the variables in the vector are various proteins and their complexes which are listed in Table 1. The classical deterministic equations constructed from these reaction network are given by,
\begin{eqnarray}
\label{deter}
\frac{dx_1}{dt}&=& -k_1x_1x_{14}+k_6-k_8x_1x_2+k_9x_4-k_{12}x_1x_6\nonumber\\
		 &&+k_{13}x_7+k_{17}x_{10}x_{12}-k_{29}x_1x_{15}\nonumber\\
		 &&+k_{33}x_{11}x_{18}+k_{34}x_9x_{15}\\
\frac{dx_2}{dt}&=& k_2x_3-k_5x_2+k_7x_4-k_8x_1x_2+k_9x_4\nonumber\\
		 &&-k_{18}x_2x_{15}-k_{20}x_2x_8+k_{21}x_{13}\nonumber\\
		 &&-k_{32}x_2x_{11}\\
\frac{dx_3}{dt}&=& k_3x_1-k_4x_3\\
\frac{dx_4}{dt}&=& -k_7x_4+k_8x_1x_2-k_9x_4-k_{19}x_4x_8\nonumber\\
		 &&-k_{30}x_4x_{15}\\
\frac{dx_5}{dt}&=& -k_{10}x_5+k_{11}x_6\\
\frac{dx_6}{dt}&=& k_{10}x_5-k_{11}x_6-k_{12}x_1x_6\\
\frac{dx_7}{dt}&=& k_{12}x_1x_6-k_{13}x_7-k_{15}x_7x_8+k_{29}x_1x_{15}\nonumber\\
		 &&+k_{31}x_{17}\\
\frac{dx_8}{dt}&=& -k_{14}x_8-k_{15}x_8x_7-k_{19}x_4x_8-k_{20}x_2x_8\nonumber\\
		 &&+k_{21}x_{13}+k_{23}-k_{35}x_8x_{15}\\
\frac{dx_9}{dt}&=& k_{15}x_8x_7-k_{16}x_9-k_{34}x_9x_{15}\\
\frac{dx_{10}}{dt}&=& k_{16}x_9-k_{17}x_{10}x_{12}-k_{33}x_{10}x_{18}\\
\frac{dx_{11}}{dt}&=& -k_{25}x_{11}x_{16}-k_{22}x_{11}+k_{24}-k_{32}x_2x_{11}\\
\frac{dx_{12}}{dt}&=& -k_{17}x_{10}x_{12}+x_{25}x_{11}x_{16}\\
\frac{dx_{13}}{dt}&=& k_{19}x_4x_8-k_{21}x_{13}\\
\frac{dx_{14}}{dt}&=& -k_1x_1x_{14}+k_{20}x_2x_8 \\
\frac{dx_{15}}{dt}&=& -k_{18}x_2x_{15}+k_{26}-k_{27}x_{15}-k_{29}x_1x_{15}\nonumber\\
		   &&-k_{30}x_4x_{15}\\
\frac{dx_{16}}{dt}&=& k_{18}x_2x_{15}-k_{25}x_{11}x_{16}-k_{28}x_{16}\nonumber\\
		   &&+k_{31}x_{17}\\
\frac{dx_{17}}{dt}&=& k_{30}x_4x_{15}-k_{31}x_{17}\\
\frac{dx_{18}}{dt}&=& k_{32}x_2x_{11}-k_{33}x_{11}x_{18}
\end{eqnarray}
where, $\{k_i\}$ and $\{x_i\}$, $i=1,2,\dots,N (N=18)$ represent the sets of rate constants of the reactions listed in Table 2 and concentration variables of the molecular species listed in Table 1. This complicated coupled set of non-linear differential equations can be solved numerically using standard fourth order Runge-Kutta method of numerical integration \cite{pre} to get the dynamical behavior of the variables listed in Table 1.

\begin{table*}
\begin{center}
{\bf Table 1 - List of molecular species} 
\begin{tabular}{|l|p{5cm}|p{6cm}|p{2.5cm}|}
 \hline \multicolumn{4}{}{} \\ \hline

\bf{ S.No.}    &   \bf{Species Name}    &    \bf{Description}           &  \bf{Notation}     \\ \hline
1.             &    $p53$               & Unbounded $p53$ protein       &  $x_1$ \\ \hline
2.             &    $Mdm2$              & Unbounded $Mdm2$ protein      &  $x_2$ \\ \hline
3.             &    $Mdm2\_mRNA$        & $Mdm2$ messenger $mRNA$       &  $x_3$  \\ \hline
4.             &    $Mdm2\_p53$         & $Mdm2$ with $p53$ complex     &  $x_4$  \\ \hline
5.             &    $ATM\_I$            & Inactivated $ATM$ protein     &  $x_5$  \\ \hline
6.             &    $ATM\_A$            & Activated $ATM$ protein       &  $x_6$  \\ \hline
7.             &    $p53\_P$            & Phosphorylated $p53$ protein  &  $x_7$  \\ \hline
8.             &    $p300$              & Unbounded $p300$ protein      &  $x_8$  \\ \hline
9.             &    $p53\_p300\_P$      & Phosphorylated $p53\_p300$ complex &  $x_9$  \\ \hline
10.            &    $p53\_A$            & Acetylated $p53$ protein      &  $x_{10}$  \\ \hline
11.            &    $HDAC1$             & Unbounded $HDAC1$ protein     &  $x_{11}$  \\ \hline
12.            &    $Mdm2\_HDAC1\_SMAR1$       & $Mdm2$, $HDAC1$ and $SMAR1$ complex    &  $x_{12}$  \\ \hline
13.            &    $Mdm2\_p53\_p300$   & $Mdm2$, $p53$ and $p300$ complex        &  $x_{13}$  \\ \hline
14.            &    $Mdm2\_p300$        & $Mdm2$ and $p300$ complex     &  $x_{14}$  \\ \hline
15.            &    $SMAR1$             &  Unbounded $SMAR1$ protein    &  $x_{15}$  \\ \hline
16.            &    $Mdm2\_SMAR1$       & $Mdm2$ and $SMAR1$ complex     &  $x_{16}$  \\ \hline
17.            &    $p53_Mdm2_SMAR1$    &  $p53$, $Mdm2$ and $SMAR1$ complex &  $x_{17}$  \\ \hline
18.            &    $HDAC1\_Mdm2$       & $HDAC1$ and $Mdm2$ complex     &  $x_{18}$  \\ \hline
\end{tabular}
\end{center}
\end{table*}
\begin{table*}
\begin{center}
{\bf Table 2 List of chemical reaction, propensity function and their rate constant} 
\begin{tabular}{|l|p{2.5cm}|p{4cm}|p{2.5cm}|p{3cm}|p{2cm}|}
\hline \multicolumn{6}{}{}\\ \hline

${\bf S.No}$ & ${\bf Reaction}$ & ${\bf Name~of~the~process}$ & ${\bf Kinetic~Law}$ & ${\bf Rate~Constant}$ & ${\bf References}$ \\ \hline
1 & $x_1+x_{14}\stackrel{k_{1}}{\longrightarrow}\phi$ & $p53$ degradation & $k_1 \langle x_1\rangle\langle x_{14}\rangle$ & $8.25\times10^{-4}{sec}^{-1}$ & \cite{gro,gro1}.\\ \hline
2 & $x_3\stackrel{k_{2}}{\longrightarrow}x_3+x_2$ & Mdm2 creation & $k_2 \langle x_3\rangle$ & $4.95\times 10^{-4}{sec}^{-1}$ & \cite{pro}.\\ \hline
3 & $x_1\stackrel{k_{3}}{\longrightarrow}x_1+x_3$ & $Mdm2\_mRNA$ creation& $k_3 \langle x_1\rangle$ & $1.0\times 10^{-4}{sec}^{-1}$ & \cite{pro}.\\ \hline
4 & $x_3\stackrel{k_{4}}{\longrightarrow}\phi$ & $Mdm2\_mRNA$ degradation & $k_4 \langle x_3\rangle$ & $1.0\times 10^{-4}{sec}^{-1}$ & \cite{pro}. \\ \hline
5 & $x_2\stackrel{k_{5}}{\longrightarrow}\phi$ & Mdm2 degradation & $k_5 \langle x_2\rangle$ & $4.33\times 10^{-4}{sec}^{-1}$ & \cite{pro}. \\ \hline
6 & $\phi\stackrel{k_{6}}{\longrightarrow}x_1$ & $p53$ synthesis & $k_6$  & $0.078{sec}^{-1}$ & \cite{pro}.\\ \hline
7 & $x_4\stackrel{k_{7}}{\longrightarrow}x_2$ & $Mdm2\_p53$ degradation & $k_7 \langle x_4\rangle$ & $8.25\times 10^{-4}{sec}^{-1}$ & \cite{mee,rod}. \\ \hline
8 & $x_1+x_2\stackrel{k_{8}}{\longrightarrow}x_4$ & $Mdm2\_p53$ synthesis & $k_8 \langle x_1\rangle\langle x_2\rangle$ & $11.55\times 10^{-4}{sec}^{-1}$ & \cite{pro}. \\ \hline
9 & $x_4\stackrel{k_{9}}{\longrightarrow}x_1+x_2$ & $Mdm2\_p53$ dissociation & $k_9 \langle x_4\rangle$ & $11.55\times 10^{-6}{sec}^{-1}$ & \cite{pro}. \\ \hline
10 & $x_5\stackrel{k_{10}}{\longrightarrow}x_6$ & ATM activation & $k_{10} \langle x_5\rangle$ & $1.0\times 10^{-4}{sec}^{-1}$ & \cite{rod,les}. \\ \hline
11 & $x_6\stackrel{k_{11}}{\longrightarrow}x_5$ & ATM deactivation & $k_{11} \langle x_6\rangle$ & $5.0\times 10^{-4}{sec}^{-1}$ & \cite{rod,les}. \\ \hline
12 & $x_1+x_6\stackrel{k_{12}}{\longrightarrow}x_6+x_7$ & Phosphorylation of $p53$ & $k_{12} \langle x_1\rangle\langle x_6\rangle$ & $5.0\times 10^{-4}{sec}^{-1}$ & \cite{rod}. \\ \hline
13 & $x_7\stackrel{k_{13}}{\longrightarrow}x_1$ & Dephosphorylation of $p53$ & $k_{13} \langle x_7\rangle$ & $5.0\times 10^{-1}{sec}^{-1}$ & \cite{rod,les}.\\ \hline
14 & $x_8\stackrel{k_{14}}{\longrightarrow}\phi$ & p300 degradation & $k_{14} \langle x_8\rangle$ & $1.0\times 10^{-4}{sec}^{-1}$ & \cite{kni,luo}. \\ \hline
15 & $x_7+x_8\stackrel{k_{15}}{\longrightarrow}x_9$ & $p53\_p300 formation$ & $k_{15} \langle x_7\rangle\langle x_8\rangle$ & $1.0\times 10^{-4}{sec}^{-1}$ & \cite{kob}. \\ \hline
16 & $x_9\stackrel{k_{16}}{\longrightarrow}x_{10}$ & Acetylation of $p53$ & $k_{16} \langle x_9\rangle$ & $1.0\times 10^{-4}{sec}^{-1}$ & \cite{gu,ito}.\\ \hline
17 & $x_{10}+x_{12}\stackrel{k_{17}}{\longrightarrow}x_1$ & Deacetylation of $p53$ & $k_{17} \langle x_{10}\rangle\langle x_{12}\rangle$ & $1.0\times 10^{-5}{sec}^{-1}$ & \cite{ito}.\\ \hline
18 & $x_{2}+x_{15}\stackrel{k_{18}}{\longrightarrow}x_{16}$ & Creation of $Mdm2\_SMAR1$ & $k_{18} \langle x_{2}\rangle\langle x_{11}\rangle$ & $2.0\times 10^{-4}{sec}^{-1}$ & \cite{ito}.\\ \hline
19 & $x_{4}+x_{8}\stackrel{k_{19}}{\longrightarrow}x_{13}$ & Creation of $Mdm2\_p53\_p300$ & $k_{19} \langle x_{4}\rangle\langle x_{8}\rangle$ & $5.0\times 10^{-4}{sec}^{-1}$ & \cite{kob}. \\ \hline
20 & $x_{2}+x_{8}\stackrel{k_{20}}{\longrightarrow}x_{14}$ & Formation of $Mdm2\_p300$ & $k_{20} \langle x_{2}\rangle\langle x_{8}\rangle$ & $5.0\times 10^{-4}{sec}^{-1}$ & \cite{pro,gro}. \\ \hline
21 & $x_{13}\stackrel{k_{21}}{\longrightarrow}x_{4}+x_{8}$ & Dissociation of $Mdm2\_$p53$\_p300$ & $k_{21} \langle x_{13}\rangle$ & $1.0\times 10^{-4}{sec}^{-1}$ & \cite{kob,mee}.\\ \hline
\end{tabular}
\end{center}
\end{table*}
\begin{table*}
\begin{center}
{\bf Continued Table 2} 
\begin{tabular}{|l|p{2.5cm}|p{4cm}|p{2.5cm}|p{3cm}|p{2cm}|}
\hline \multicolumn{6}{}{}\\ \hline
22 & $x_{11}\stackrel{k_{22}}{\longrightarrow}\phi$ & Degradation of HDAC1 & $k_{22} \langle x_{11}\rangle$ & $1.0\times 10^{-4}{sec}^{-1}$ & \cite{ito}. \\ \hline
23 & $\phi\stackrel{k_{23}}{\longrightarrow}x_{8}$ & p300 synthesis & $k_{23} ~(k_{p300})$ & $0.1{sec}^{-1}$ & \cite{kni,luo}. \\ \hline
24 & $\phi\stackrel{k_{24}}{\longrightarrow}x_{11}$ & HDAC1 synthesis & $k_{24}~(k_{HDAC1})$ & $2.0\times 10^{-4}{sec}^{-1}$ & \cite{ito}.\\ \hline
25 & $x_{11}+x_{16}\stackrel{k_{25}}{\longrightarrow}x_{12}$ & Synthesis of $SMAR1\_Mdm2\_HDAC1$ & $k_{25} \langle x_{11}\rangle$ & $1.0\times 10^{-4}{sec}^{-1}$ & \cite{ito}. \\ \hline
26 & $\phi\stackrel{k_{26}}{\longrightarrow}x_{15}$ & $SMAR1$ synthesis & $k_{26}$ & $0.08{sec}^{-1}$ & \cite{kni,luo}. \\ \hline
27 & $x_{15}\stackrel{k_{27}}{\longrightarrow}\phi$ & Degradation of $SMAR1$ & $k_{27}$ & $1.0\times 10^{-4}{sec}^{-1}$ & \cite{ito}.\\ \hline
28 & $x_{16}\stackrel{k_{28}}{\longrightarrow}\phi$ & Degradation of $SMAR1\_Mdm2$ complex & $k_{28} \langle x_{11}\rangle$ & $2.0\times 10^{-4}{sec}^{-1}$ & \cite{ito}. \\ \hline
29 & $x_1+x_{15}\stackrel{k_{29}}{\longrightarrow}x_7$ & Phosphorylation of $p53$ & $k_{26} \langle x_{1}\rangle\langle x_{15}\rangle$ & $1.0\times 10^{-4}{sec}^{-1}$ & \cite{kni,luo}. \\ \hline
30 & $x_4+x_{15}\stackrel{k_{30}}{\longrightarrow}x_{17}$ & Synthesis of $p53\_Mdm2\_SMAR1$ complex & $k_{30}\langle x_4\rangle\langle x_{15}\rangle$ & $1.0\times 10^{-3}{sec}^{-1}$ & \cite{ito}.\\ \hline
31 & $x_{17}\stackrel{k_{31}}{\longrightarrow}x_7+x_{16}$ & Phosphorylation of $p53$ & $k_{31} \langle x_{17}\rangle$ & $1.0\times 10^{-3}{sec}^{-1}$ & \cite{ito}. \\ \hline
32 & $x_2+x_{11}\stackrel{k_{32}}{\longrightarrow}x_{18}$ & $Mdm2$\_HDAC1 complex synthesis & $k_{32}\langle x_2\rangle\langle x_{11}\rangle$ & $2.0\times 10^{-3}{sec}^{-1}$ & \cite{ito}.\\ \hline
33 & $x_{18}+x_{10}\stackrel{k_{33}}{\longrightarrow}x_1$ & Synthesis of $p53$ & $k_{33} \langle x_{18}\rangle\langle x_{10}\rangle$ & $5.0{sec}^{-1}$ & \cite{kni,luo}. \\ \hline
34 & $x_9+x_{15}\stackrel{k_{34}}{\longrightarrow}x_1+x_{15}$ & Interaction of $p53\_p300$ complex with $SMAR1$ & $k_{34}\langle x_9\rangle\langle x_{15}\rangle$ & $1.0\times 10^{-4}{sec}^{-1}$ & \cite{ito}.\\ \hline
35 & $x_8+x_{15}\stackrel{k_{35}}{\longrightarrow}x_{15}$ & Degradation of p300 by $SMAR1$ & $k_{35} \langle x_8\rangle\langle x_{15}\rangle$ & $5.0\times 10^{-1}{sec}^{-1}$ & \cite{kni,luo}. \\ \hline
\end{tabular}
\end{center}
\end{table*}

\begin{figure}
\label{schm}
\begin{center}
\includegraphics[height=8cm,width=8.5cm]{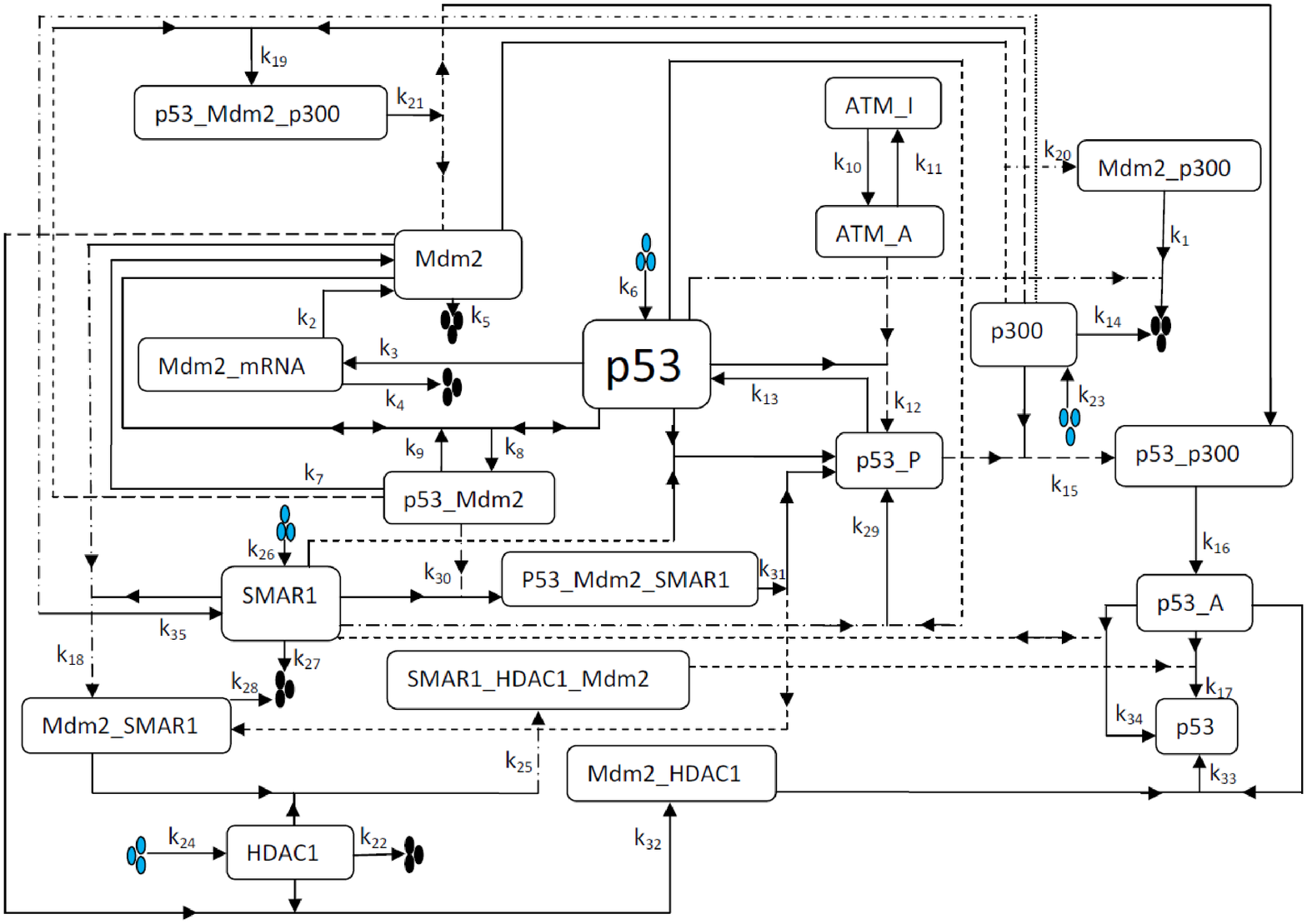}
\caption{The schematic diagram of $p53-Mdm2-p300-HDAC1-SMAR1$ regulatory network.} 
\end{center}
\end{figure}

\begin{figure}
\label{fig2}
\begin{center}
\includegraphics[height=8.0cm,width=8.5cm]{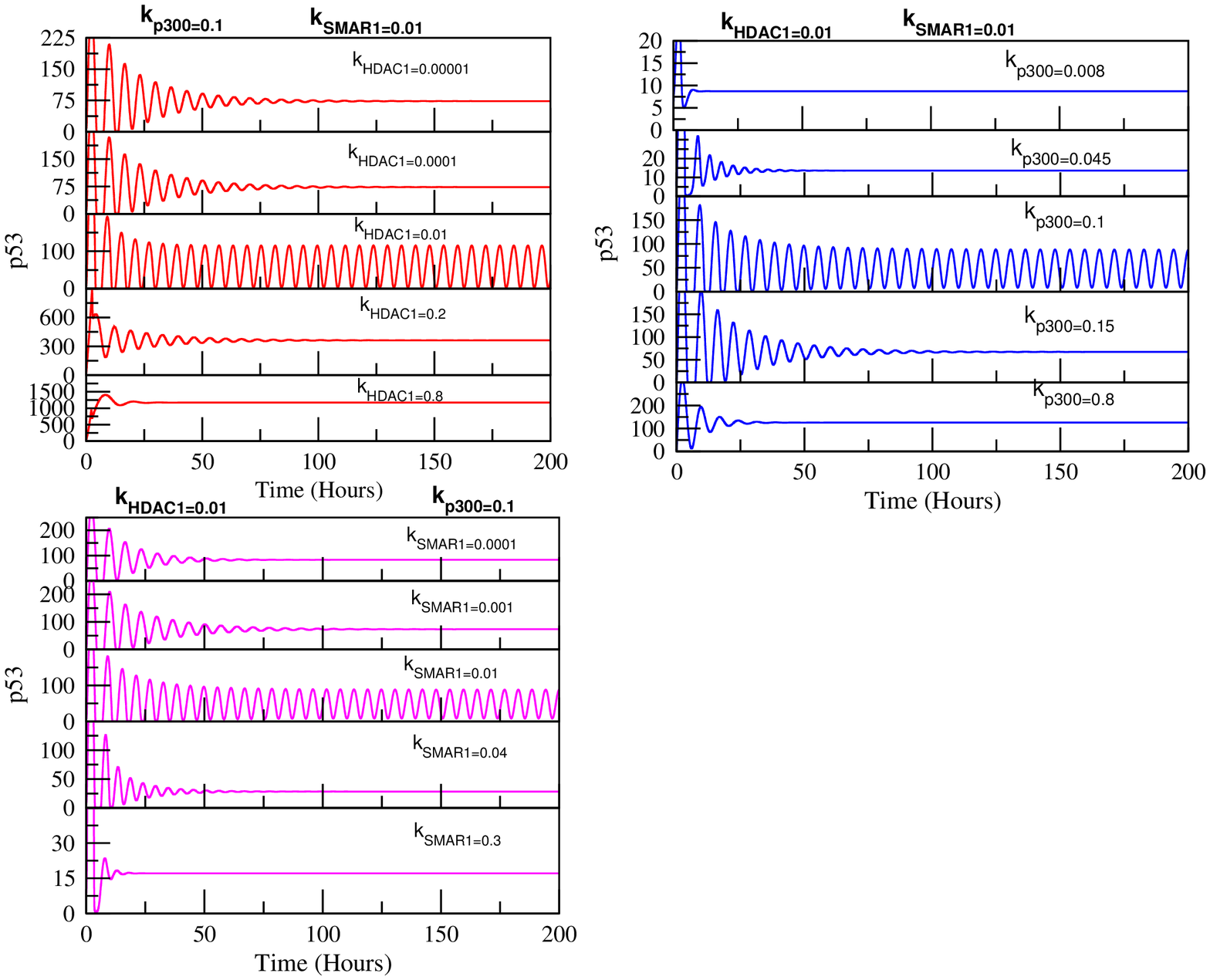}
\vskip 0.7cm
\includegraphics[height=2.0cm,width=8.5cm]{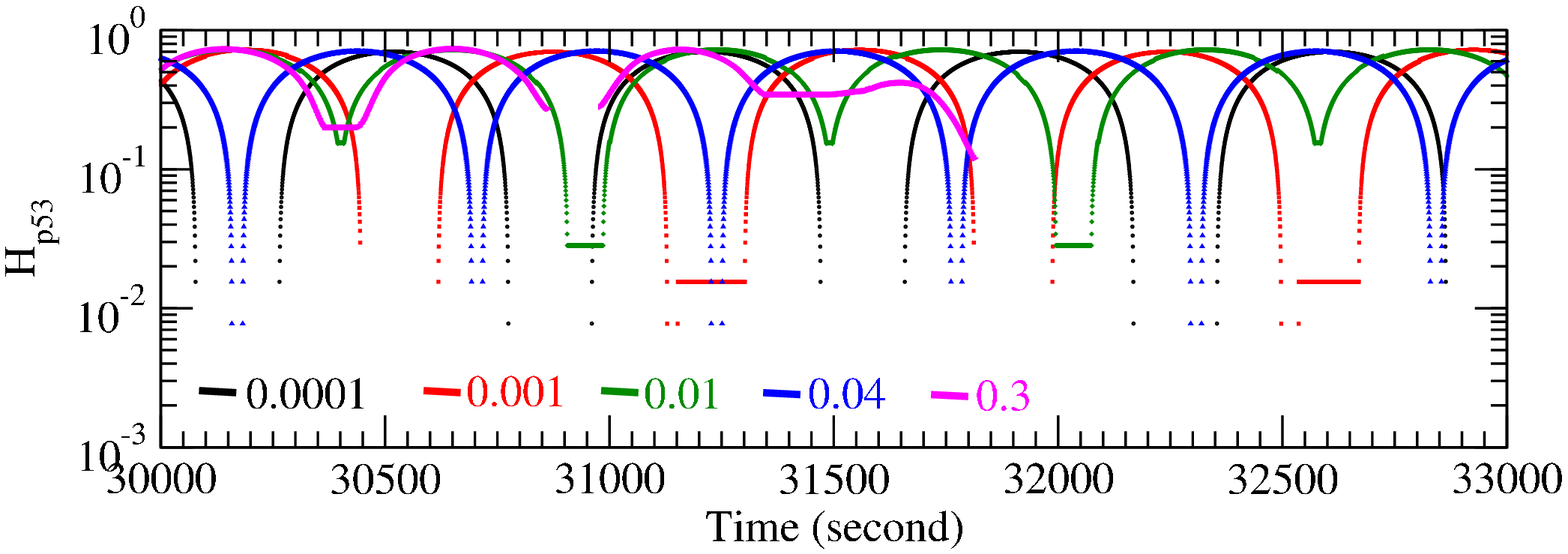}
\caption{(A) The dynamics of $p53$ induced by $k_{SMAR1}$, $k_{p300}$ and $k_{HDAC1}$ showing three different distinct states, namely, stabilized (2), damped and sustain oscillation states in all three cases. (B) The permutation entropy spectrum of the three states of $p53$ dynamics induced by $SMAR1$.} 
\end{center}
\end{figure}

Fluctuations in $p53$ network dynamics triggered by stress inducing molecular species could highlight some of basic regulatory mechanisms of how regulatory network works and self-organized by itself to maintain normal functioning of the network. The $p53$ network sees a stress inducing molecular species not as a single species but as a sub-network of that species in which the species itself is the central node (removing this node cause break down of the sub-network). Fluctuations in any one of the reaction channel in the sub-network cause changes in all the components of the sub-network, and then impart that overall perturbation to the main $p53$ network which alters the topological characteristics and dynamics of the network. These changes in the properties of the network induce fluctuations in the properties of individual behavior of the components of the network.

The state of the dynamical system given by coupled ordinary differential equations (ODE) (1)-(18) at any instant of time 't' is given by state vector, ${\vec x}(t)=(x_1,x_2,...,x_N)^T$, where, $T$ is the transpose of the vector and $N=18$. The system of reactions (Table 2), from which the ODEs (1)-(18) are constructed, can be approximately divided into two types of elementary reactions, namely \textit{fast} and \textit{slow} reactions \cite{murr}. The variables in the state vector $\vec{x}$ can be divided into fast and slow vectors given by,
\begin{eqnarray}
\label{steady}
\vec{x}^s=\left[\begin{matrix}x_1\\x_2\\x_8\\x_{11}\\x_{15}\end{matrix}\right];~
\vec{x}^f=\left[\begin{matrix}x_3\\x_4\\x_5\\x_{6}\\x_{7}\\x_9\\x_{10}\\x_{12}\\x_{13}\\x_{14}\\x_{16}\\x_{17}\\x_{18}\end{matrix}\right];~
\vec{x}=\left[\begin{matrix}\vec{x}^s\\\vec{x}^f  \end{matrix}\right]
\end{eqnarray}
The fast variables are normally corresponding to complex molecular species. Generally, formation of complex molecular species due to fast reactions are followed by fast decay of these complexes, the dynamics of the fast variables reach steady state much quickly as compared to the dynamics of slow variables \cite{scha,pede}. We then use Henri-Michaelis-Menten-Briggs-Haldane approximation to assume that the time evolution of fast state vector $\vec{x}^f$ reach equilibrium state defined by $\vec{x}^{*f}$ much faster as compared to the time evolution of slow state vector $\vec{x}^s$ \cite{scha,pede}. Applying this approximation, we can reach the following steady state for fast variables,
\begin{eqnarray}
\label{fix}
\frac{d\vec{x}^f}{dt}\approx 0;~\vec{x}^{*f}=\left[\begin{matrix}x_3^*\\x_4^*\\x_5^*\\x_{6}^*\\x_{7}^*\\x_9^*\\x_{10}^*\\x_{12}^*\\x_{13}^*\\x_{14}^*\\x_{16}^*\\x_{17}^*\\x_{18}^*\end{matrix}\right]
\end{eqnarray}
such that the dynamics of the system for sufficiently large time is governed by the dynamics of the slow variables given by,
\begin{eqnarray}
\label{quasi}
\frac{d\vec{x}}{dt}=\frac{d}{dt}\left[\begin{matrix}\vec{x}^s\\\vec{x}^f\end{matrix}\right]\approx\frac{d\vec{x}^s}{dt}=\frac{d}{dt}\left[\begin{matrix}x_1\\x_2\\x_8\\x_{11}\\x_{15}\end{matrix}\right]
\end{eqnarray}
The approximate solution of the complex model can be obtained from this reduced model using quasi steady state approximation.

\subsection{Permutation entropy: measure of complexity}
The information contain in a system is reflected in the complexity in time series of the constituting variables of the system which can be quantified by the measure of permutation entropy of the time series \cite{band,cao}. The permutation entropy $H$ of a time series of a constituting variable $x(t)$ of a system can be calculated as follows. The time series of the variable $x(t)$ can be mapped onto a symbolic sequence of length $N$: $x(t)=\{x_1,x_2,...,x_N\}$. This sequence is then partitioned into $M$ number of short sequences of length $U$ which can be expressed as $x(t)=\{w_1,w_2,...,w_M\}$, where ith window is given by, $w_i=\{x_{i+1},x_{i+2},...,x_{i+U}\}$ which allows to slide along the sequence $x(t)$ with maximum overlap. To calculate permutation entropy of a window $w_i$, one defines a short sequence of embedded dimension $r$, $S_i=\{x_{i+1},x_{i+2},...,x_{i+r}\}$ in r-dimensional space, finding all possible inequalities of dimension $r$ and mapping the inequalities along the ascendingly arranged elements of $w_i$ to find the probabilities of occurrence of each inequality in $w_i$. Since $q$ out of $r!$ permutations are distinct, one can define a normalized permutation entropy as $H_i=-\frac{1}{ln(r!)}\sum_{j=1}^{q}p_jln(p_j)$ where, $0\le H_i(r)\le 1$. The mapped permutation entropy spectrum of time series $x(t)$ is $H=\{H_1,H_2,...,H_M\}$ which is the measure of complexity of time series $x(t)$. For self-organized state corresponds to order state giving $H\rightarrow 0$.

\section{Results and Discussion}
Based on above equation we have obtained from our proposed integrated model, the simulation have been done. Here we only limits our study upto the deterministic solution of the equation. We have solved the set of differential equation using standard runge-kutte 4th order differential equation. 

\subsection{Approximate solution of the model}
The fast state vector reach steady state quickly and can be taken as constant as compared to slow state variable (equations (\ref{fix}) and (\ref{quasi})). From equations (18) and (\ref{fix}), one can reach $x_2=\frac{k_{33}}{k_{32}}x_{18}^*$ showing the direct dependence of $x_2$ on $x_{18}^*$ ie steady state of HDAC1\_$Mdm2$ complex. Similarly, from equations (3) and (\ref{fix}) we get $x_1=\frac{k_4}{k_3}x_3^*$ indicating direct proportional to the steady state of $Mdm2$\_mRNA complex. Putting these equations to equation (15) and using equation (\ref{fix}), we have the following equation,
\begin{eqnarray}
\label{dx15}
\frac{dx_{15}}{dt}+Ux_{15}=k_{26}
\end{eqnarray}
where, $U=k_{27}+\frac{k_{18}k_{33}}{k_{32}}x_{18}^*+\frac{k_4k_{29}}{k_3}x_3^*+k_{30}x_4^*$ is a constant within quasi-steady state approximation. The solution of this equation (\ref{dx15}) is given by,
\begin{eqnarray}
\label{x15}
x_{15}(t)=\frac{k_{26}}{U}\left(1-e^{-Ut}\right)+x_{15}(0)e^{-Ut}
\end{eqnarray}
where, $x_{15}(0)$ is the initial concentration of $x_{15}$ at $t=0$. The solution (\ref{x15}) shows that the rate of increase of $x_{15}$ $SMAR1$ in the system is restricted by the steady state values of $x_3^*$, $x_{4}$ and $x_{18}$ via $U$; and time. The asymptotic value of $x_{15}$ as $t\rightarrow\infty$ is found to be $x_{15}\approx\frac{k_{26}}{U}$ reaching a steady state. For small values of time 't', keeping upto linear terms in the expansion $e^{xt}\sim 1+xt+O(t^2)$, we get $x_{15}(t)\approx\left[k_{26}-Ux_{15}(0)\right]t$ which shows minimal sufficient condition for $x_{15}$ creation is $k_{26}>Ux_{15}(0)$.
\begin{figure}
\label{fig3}
\begin{center}
\includegraphics[height=8.5cm,width=8.5cm]{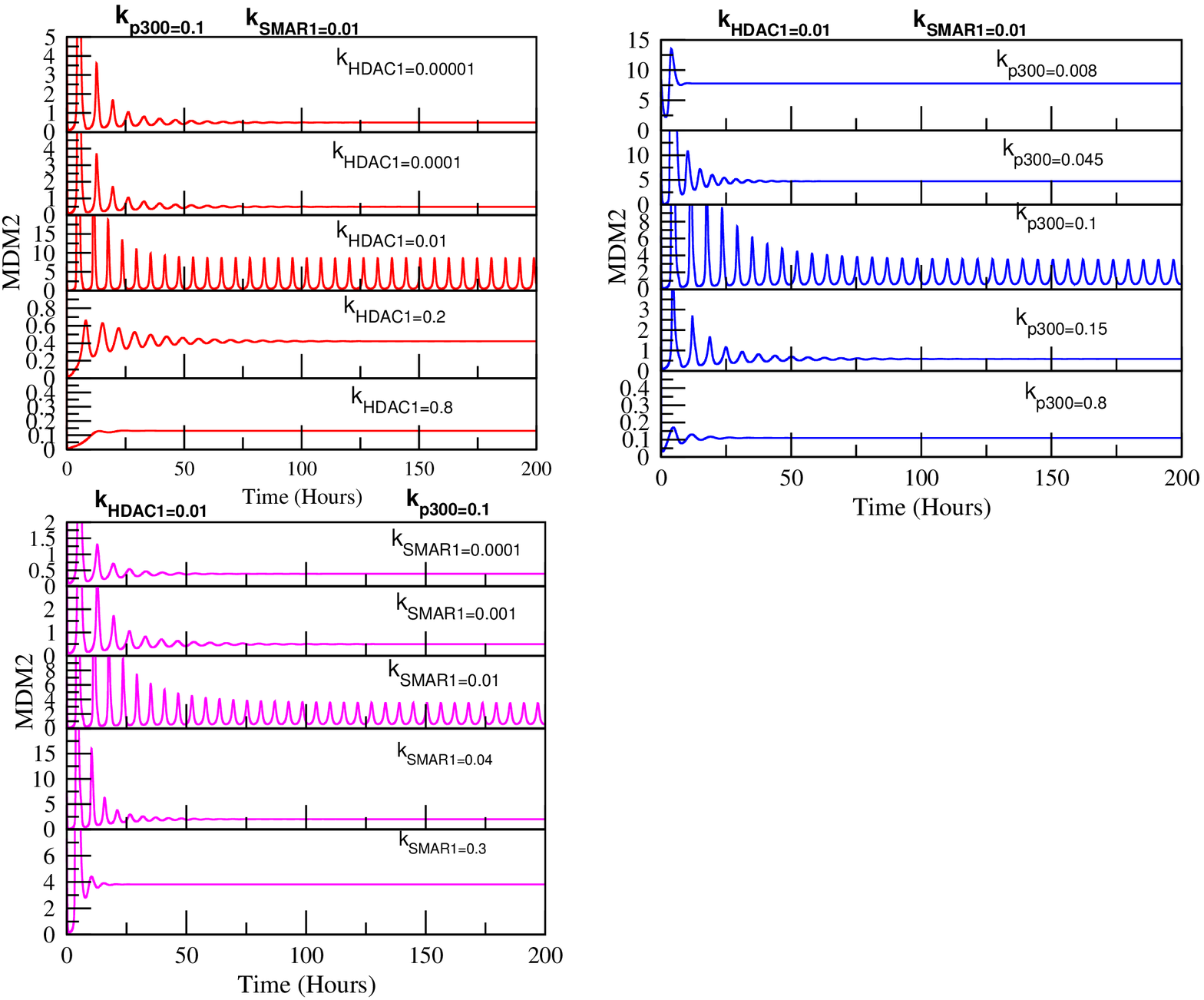}
\vskip 0.7cm
\includegraphics[height=2.0cm,width=8.5cm]{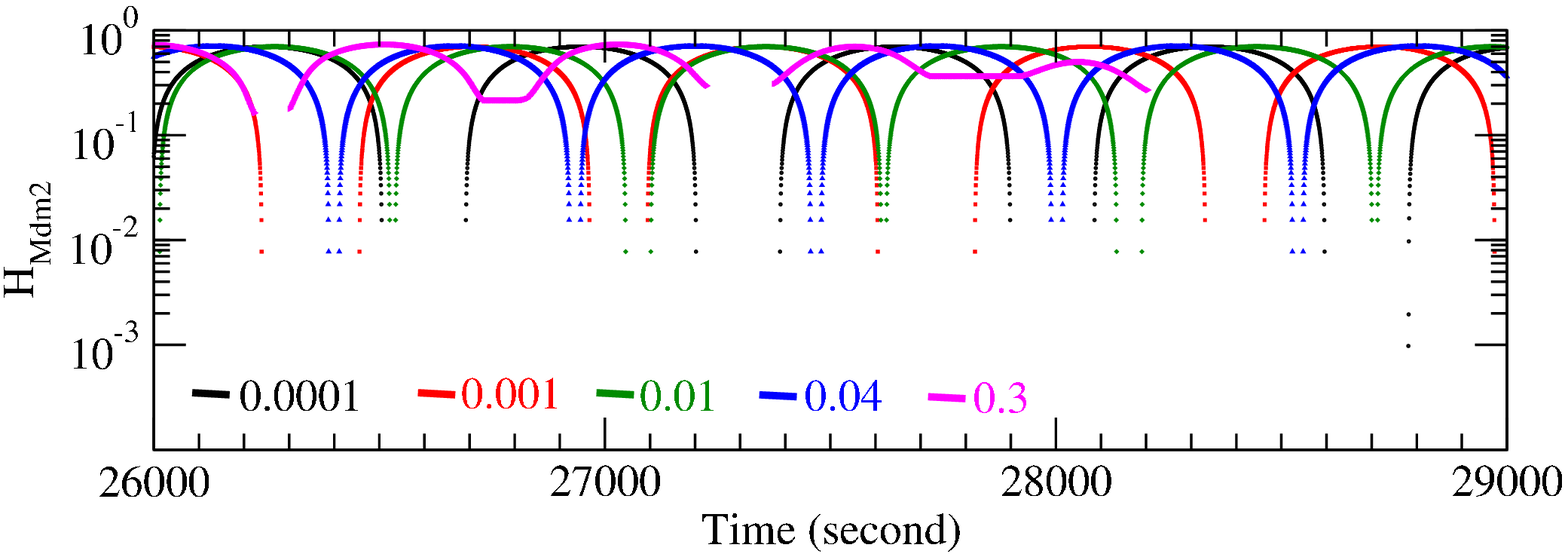}
\caption{(A) The $Mdm2$ dynamics induced by stress inducers, $k_{SMAR1}$, $k_{p300}$ and $k_{HDAC1}$ showing three different distinct states as obtained in the case of $p53$. (B) The permutation entropy spectrum of the three states of $Mdm2$ dynamics induced by $SMAR1$.} 
\end{center}
\end{figure}

The equations (11), (\ref{quasi}), (\ref{fix}) and $x_2=\frac{k_{33}}{k_{32}}x_{18}^*$ can be used to get the following ODE of variable $x_{11}$,
\begin{eqnarray}
\label{dx11}
\frac{dx_{11}}{dt}+Vx_{11}=k_{24}
\end{eqnarray}
where, $V=k_{22}+k_{25}x_{16}^*+k_{33}x_{18}^*$ is a constant. Then the solution of the ODE (\ref{dx11}) can be obtained given by,
\begin{eqnarray}
\label{x11}
x_{11}(t)=\frac{k_{24}}{V}\left(1-e^{-Vt}\right)+x_{11}(0)e^{-Vt}
\end{eqnarray}
where, $x_{11}(0)$ is the initial value of $x_{11}$ at $t=0$. The asymptotic value of $x_{11}$ at $t\rightarrow\infty$ is given by $x_{11}\approx\frac{k_{24}}{V}$ which is the steady state. At small time limit where exponential expansion is approximated upto linear terms, we obtain $x_{11}(t)\sim\left[k_{24}-Vx_{11}(0)\right]t$. The minimal sufficient condition for formation of $x_{11}$ is $k_{24}>Vx_{11}(0)$.

Similarly, using equations (8), (\ref{quasi}) and (\ref{fix}) we can reach the following ODE,
\begin{eqnarray}
\label{dx8}
&&\frac{dx_{8}}{dt}+x_{8}\left[W+k_{35}\left(\frac{k_{26}}{U}+We^{-Ut}\right)\right]\nonumber\\
&&=k_{23}+k_{13}k_{21}
\end{eqnarray}
where, $W=k_{14}+k_{15}x_7^*+k_{19}x_4^*+\frac{k_{20}k_{33}}{k_{32}}x_{18}^*$ is a constant. The solution of this ODE can be obtained by taking $\int\rightarrow\int_{0}^{\infty}$ which is also true for positive values of $x_{8}$, and is given by,
\begin{eqnarray}
\label{x8}
x_{8}(t)=\left[C_1-G\left(\frac{U}{W}\right)^{H/U}\Gamma\left(\frac{H}{U}\right)\right]e^{-Ht+\frac{W}{U}e^{-Ut}}
\end{eqnarray}
where, $G=\frac{k_{23}+k_{21}x_{18}^*}{U}$ and $H=W+\frac{k_{26}k_{35}}{U}$ are constants. The constant $C_1$ can be obtained by using initial condition i.e. $t=0$. Putting back the expression for $C_1$ to equation (\ref{x8}), we get,
\begin{eqnarray}
\label{xx8}
x_{8}(t)=x_8(0)e^{-\frac{W}{U}}e^{-Ht+\frac{W}{U}e^{-Ut}}
\end{eqnarray}
It is observed that for large value of $t$, the term $Ht$ dominates $e^{-Ut}$, and therefore we have $x_{8}(t)\propto e^{-Ht}$. However, for small 't', we have $x_8(t)\sim x_8(0)e^{-W/U}\left[1-(H+W)t\right]$, which indicates that the minimal existence of $x_{8}$ will have the condition $(H+W)t<1$.

Now, to get the solution for $x_2$, the equations (12) and (16) using (\ref{fix}) are added, and the result is substituated in equation (2). The simplified ODE of $x_{2}$ is given by,
\begin{eqnarray}
\label{dx2}
\frac{dx_{2}}{dt}+x_{2}\left[R+\frac{S}{e^{Vt}}\right]=D
\end{eqnarray}
where, $R=k_5+\frac{k_{24}k_{32}}{V}$, $S=k_{32}\left(x_{11}-\frac{k_{24}}{V}\right)$ and $D=k_2x_3^*+k_7x_4^*+k_{21}x_{13}^*+k_{31}x_{17}^*-\frac{k_4x_8}{k_3}x_3^*-k_{17}x_{10}^*x_{12}^*-k_{28}x_{16}^*-\frac{k_1k_4}{k_3}x_3^*$ are constants. The solution of the equation (\ref{dx2}) is given by,
\begin{eqnarray}
\label{x2}
x_{2}(t)=\left[C_2-\frac{D}{S}\left(\frac{V}{S}\right)^{\frac{R}{V}-1}\Gamma\left(\frac{R}{V}\right)\right]e^{-Rt+\frac{S}{V}e^{-Vt}}
\end{eqnarray}
where, $C_2$ is a constant which can be obtained from initial condition $t=0$, Then putting back the expression for $C_2$ to the equation (\ref{x2}), we get,
\begin{eqnarray}
\label{xx2}
x_{2}(t)=x_2(0)e^{-\frac{S}{V}}e^{-Rt+\frac{S}{V}e^{-Vt}}
\end{eqnarray}
The large 't' limit in the equation (\ref{xx2}) show that $x_2(t)\sim x_2(0)e^{-S/V}e^{-Rt}$ which shows that $x_2(t)\propto e^{-Rt}$. However, it further indicates that $\lim_{t\rightarrow\infty}x_2(t)=0$. Small 't' approximation to the equation (\ref{xx2}) leads to the expression $x_2(t)\sim x_2(0)e^{-S/V}\left[1-(R+S)t\right]$, which shows that minimal condition for existence of $x_2$ is $1>(R+S)t$.
\begin{figure}
\label{fig4}
\begin{center}
\includegraphics[height=8.5cm,width=8.5cm]{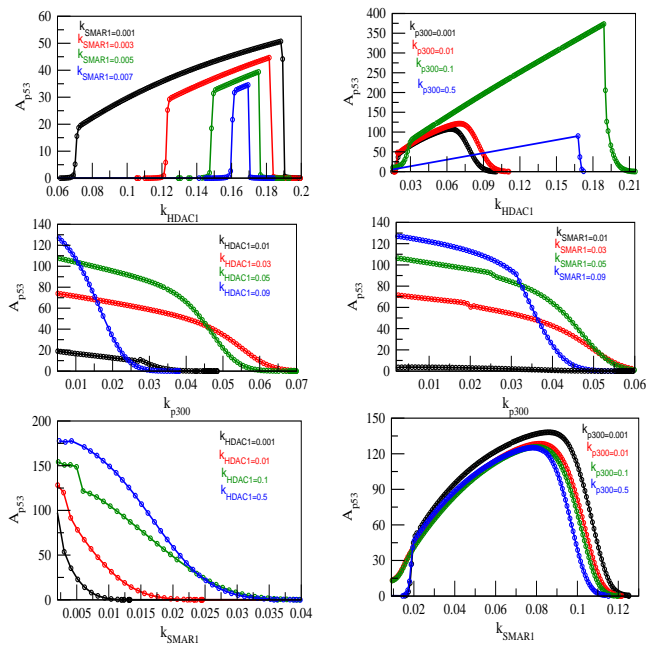}
\caption{The variation of amplitude of $p53$ dynamics induced by $k_{SMAR1}$, $k_{p300}$ and $k_{HDAC1}$ which shows three different states: stabilized state (one for normal and the other for apoptotic states, indicated by amplitude death case), damped states and sustain states.} 
\end{center}
\end{figure}

Similarly, proceeding same way as above, from equations (1), (7), (9) and (steady) we obtain the following ODE for $x_1$,
\begin{eqnarray}
\label{dx1}
\frac{dx_{1}}{dt}+Fx_{1}=G+Pe^{-Vt}
\end{eqnarray}
where, $F=k_1x_{14}^*+\frac{k_8k_{13}k_{33}}{k_{32}}$, $G=k_6+k_9x_4^*+k_{13}x_7^*+k_{17}x_{10}x_{12}^*+k_{31}x_{17}^*-k_{13}x_7^*-k_{16}x_9^*+\frac{k_{18}k_{24}k_{33}}{V}$, and $P=k_{33}k_{18}$ are constants. The solution of the equation (\ref{dx1}) is given by,
\begin{eqnarray}
\label{x1}
x_{1}(t)&=&\frac{G}{F}\left(1-e^{Ft}\right)+\frac{P}{F-V}\left(e^{-Vt}-e^{-Ft}\right)\nonumber\\
&&+x_1(0)e^{-Ft}
\end{eqnarray}
Now, the small 't' approximation allows to simplify equation (\ref{x1}) to obtain $x_1(t)\sim x_1(0)+t[P-G-Fx_1(0)]$. The minimal condition for $x_1$ existence in the system is given by, $x_1(0)>t[G+Fx_1(0)-P]$. However, in the large 't' approximation, we have, $x_{1}(t)=\frac{G}{F}\left(1-e^{Ft}\right)$ and for non-negative value of $x_1$ the condition is, $e^{Ft}<1$. However, we have $\lim_{t\rightarrow\infty}x_1(t)=-\infty$.

\subsection{Complexity in transition of $p53$ states} 
The changes in the states of the dynamics of $p53$ are triggered by various signaling molecules, namely, $SMAR1$, p300 and HDAC1 respectively (Fig. 2) and found three distinct states, two steady states, damped oscillation state and sustain oscillation states. When $p53$ regulatory network interacts with one of the signaling molecules, the network saw that signaling molecule as a sub-network (see Fig. 1) which involves a number of interaction and a number of complexes due to the interaction. So the changes in the $p53$ dynamics are due to the fluctuations in the sub-network associated with the signaling molecule.

The concentration of $HDAC1$ in the system depends on the value of creation rate of it $k_{24}$ which we have taken as $k_{HDAC1}$ in our simulation. At low concentration of $HDAC1$ (small value of $k_{HDAC1}$) allows $p53$ to maintain its normal state (stabilized state) in the system (Fig. 2 left upper panels). As one increase the concentration of $HDAC1$ in the system (increasing the value of $k_{HDAC1}$), $HDAC1$ starts active interaction with $Mdm2$ and $SMAR1$ forming various complexes followed by indirect interaction with $p53$ (Table 2). This indirect interaction of $HDAC1$ and $p53$ impart stress in $p53$ dynamics which starts exhibit damped oscillation (mixture of stress and stabilized state) indicating the induction of stress by the available $HDAC1$ concentration in the system and then come back to the normal state\cite{ito1,luo}. The range of damped oscillation increases as $k_{HDAC1}$ value increases and after sufficient value of $k_{HDAC1}$, $p53$ dynamics become sustain oscillation for a certain range of $k_{HDAC1}\rightarrow [0.007-0.05]$. This sustain oscillation state corresponds to strong activated or stress state which is found to be maximum at $k_{HDAC1}=0.07$ (where amplitude of $p53$ of the corresponding sustain oscillation is maximum$\sim 123.2\pm 2$), then start decreasing as $k_{HDAC1}$ increases. After $k_{HDAC1}>0.05$, the dynamics of $p53$ become damped oscillation, which indicates that large $HDAC1$ concentration in the system trigger large stress which can't be repair back and may probably go to apoptosis. This range of stress in this case decreases with amplitude as the value of $k_{HDAC1}$ increases. Excess $HDAC1$ concentration in the system may trigger immediate apoptosis of the system (stabilized state)\cite{luo1,bol,ale}.

Similarly, the three states of $p53$ are also found when the $p53$ regulatory network is perturbed by sub-network of $SMAR1$ (Fig. 2 middle panels) which is composed of $SMAR1$ and its interaction partners i.e. associated complexes (Fig. 1) and acts as main hub in the sub-network. The rate constant of formation of $SMAR1$ in the system $k_{26}$, which we take $k_{SMAR1}$ as notation, corresponds to the availability of $SMAR1$ concentration in the system to induce perturbation in $p53$ network\cite{sin}. This accessible concentration of $SMAR1$ affects the dynamics its own sub-network, and then impart perturbation to $p53$ network. The results of perturbation, similar to that of $HDAC1$, shows nearly normal state for $k_{SMAR1}<0.0001$ for fixed values of $k_{HDAC1}=0.01$ and $k_{p300}=0.1$, damped states in two ranges $[0.0001-0.005]$ (increasing range of damped oscillation as $k_{SMAR1}$ increases) and $[0.06-0.28]$ (decreasing range of damped oscillation as $k_{SMAR1}$ increases), and sustain oscillation in the range $[0.0051-0.058]$ which decreases $p53$ amplitude as $k_{SMAR1}$ increases. Therefore, excess $SMAR1$ concentration in the system triggers apoptosis. Similar behavior is found in $Mdm2$ case (Fig. 3).

Similar behavior of these three states is found for the case of $p300$ induced $p53$ dynamics (Fig. 2 right panels). Similar behavior is found in $Mdm2$ case (Fig. 3). This reveals that this signaling molecule has also the tendency to induce apoptosis in the system \cite{aro, ito}.

The permutation entropies $H_{p53}$ of the three states of $p53$ driven by $SMAR1$ are alculated for $p53$ dynamics to understand complexity of the perturbed network (Fig. 2 lowermost panel). We took embedded dimension $r=3$ and window size to be $w_s=512$. We also tried for other values of embedded dimension i.e. 4, 5 and 6, and found the results almost the same. The results show that for normal state (low value of $k_{SMAR1}=0.0001$) the values of $H_{p53}$ is low, with large gaps among nearly periodic curves which consist of large number of near zero points. This low values of $H_{p53}$ indicates more self-organized behavior at normal state of the system. If we increase the values of $k_{SMAR1}$ ($k_{SMAR1}=0.001,0.01,0.04$) the $H_{p53}$ values start increasing, and the gap between neighbouring curves decreases, showing significant increase of $H_{p53}$ points as compared to normal state. This indicates that increase in stress due to increase in $SMAR1$ concentration in the system disturb the self-organized behavior forcing the system to more complex state. The second stabilized state (with excess $SMAR1$ concentration in the system corresponding to $k_{SMAR1}=0.3$) or apoptotic state has maximum $H_{p53}$ value showing most disorganized state of the system. Similar behavior is found for $Mdm2$ case (Fig. 3 lowermost panel)\cite{biz}.
\begin{figure}
\label{fig5}
\begin{center}
\includegraphics[height=8.5cm,width=8.5cm]{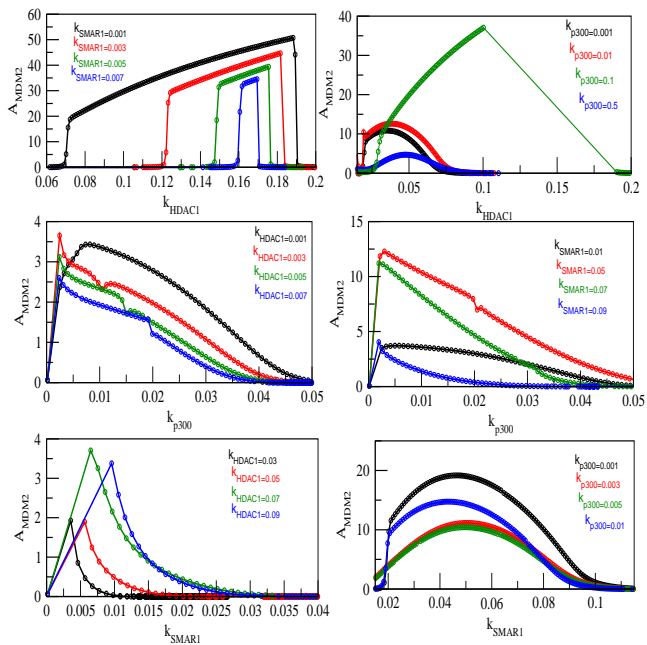}
\caption{The variation of amplitude of $Mdm2$ dynamics induced by $k_{SMAR1}$, $k_{p300}$ and $k_{HDAC1}$ which shows three different states: stabilized state (one for normal and the other for apoptotic states, indicated by amplitude death case), damped states and sustain states.} 
\end{center}
\end{figure}

\subsection{Amplitude death: signature of apoptosis}
The amplitude of $p53$ oscillatory dynamics due to fluctuations induced by changes in some part of the network (for example changes in concentration of $SMAR1$, $p300$, $HDAC1$ with corresponding sub-networks associated with them) refers to the amount of stress induced in its dynamics. The amount of stress imparted in the system allows active interaction of this $p53$ with the respective fluctuated molecular species directly or indirectly, and once the stress is removed, the active interaction stays for sometime with damped oscillation which we call $"Relaxation~time"$ and come back to normal situation where the amplitude becomes zero (amplitude death) (Fig. 2 and Fig. 3). The relaxation time increases as the amount of stress is increased, and become infinite for certain range of value of stress parameter ($k_{SMAR1}$, $k_{p300}$, $k_{HDAC1}$ etc) which is the case of sustain oscillation. Larger values of stress parameter than this range, the transition of sustain to damped oscillation states takes place. Further larger values of stress parameter force the dynamics to amplitude death scenario again (Fig. 4 and Fig. 5). This transition of various oscillating states as a function of stress parameter give corresponding signatures of the state of the system\cite{ala}.

Normal state of $p53$ dynamics (small values of stress parameter) show amplitude death scenario of $p53$ as a function of $k_{HDAC1}$ for different values of $k_{SMAR1}$ (Fig. 4 upper left panel). The transition from amplitude death (normal state) to damped state (mixture of stress then come back to normal after removing of stress) is for small range of $k_{HDAC1}$ only, and suddenly move to the sustain oscillation state (we took long time series of 500 hours i.e. 5 days duration after removing transients). Within the range of sustain oscillation state, the amplitude of $p53$ ($A_{p53}$) increases as a function of $k_{HDAC1}$ (equation (\ref{x1})). The amplitude $A_{p53}$ suddenly drops to zero (amplitude death scenario) after a short range $k_{HDAC1}$. This second regime of amplitude death scenario could be the apoptotic state of the modeled system. Further, it can also be seen that for the same range of $k_{HDAC1}$, as $k_{SMAR1}$ increases the range of sustain oscillation decreases and on the other hand the amplitude of $p53$ decreases. It reveals that if the value of $k_{SMAR1}$ is large enough the the system will go to amplitude death (apoptotic) regime directly. Similar transition the states can also be found in the case of $Mdm2$ dynamics also (Fig. 5).

The transition of the states can also be seen in the parameter space of $p53$ and $k_{HDAC1}$ for different values of $k_{p300}$ and for fixed value of $k_{SMAR1}$ (Fig. 4 upper right panel). The different in behavior in this is the increase in the regime of sustain oscillation as $k_{p300}$ increases until $k_{p300}=0.1$. This indicates that within this range of $k_{p300}$, increasing $k_{p300}$ can able to increase the range of $k_{HDAC1}$ before reaching apoptosis. After this value the $A_{p53}$ behavior does not usual transition and goes to zero amplitude quickly (Fig. 4 upper right panel). This means that one can engineer the modeled system in such a way that increasing $k_{p300}$ can able to increase the accessible $k_{HDAC1}$ to save the system from apoptosis.
\begin{figure}
\label{fig6}
\begin{center}
\includegraphics[height=12.0cm,width=6.0cm]{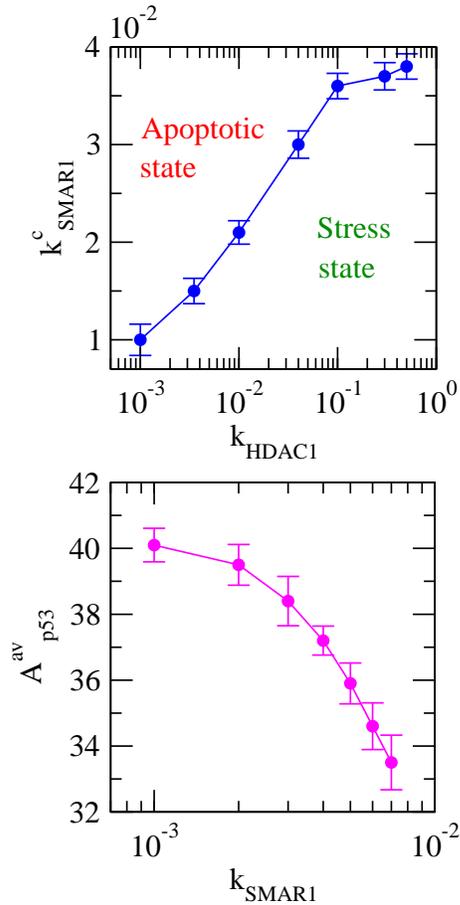}
\caption{(A) Phase diagram of $p53$ dynamics in the parameter space ($k_{HDAC1},k^c_{SMAR1}$) showing distinct phases, namely, stress and apoptotic phases. (B) The plot of $A^{av}_{p53}$ as a function of $k_{SMAR1}$.} 
\end{center}
\end{figure}
 
The behavior of $A_{p53}$ as a function of $k_{p300}$ for various values of $k_{HDAC1}$ and for a fixed value of $k_{SMAR1}$ shows two states transition, namely sustain oscillation and amplitude death (apoptotic state) (Fig. 4 left middle panel). This is due to the choice of value of $k_{SMAR1}$ is to induce sustain oscillation. The increase in $k_{HDAC1}$ allow the system to reach amplitude death regime quickly. Same is true for the case of $A_{53}$ versus $k_{p300}$ for various values of $k_{SMAR1}$ (Fig. 4 right middle panel).

The scenario of transition of the states is different in the case of $A_{p53}$ as a function of $k_{SMAR1}$ for various values of $k_{HDAC1}$ which shows the increase in the range of accessible $k_{SMAR1}$ as $k_{HDAC1}$ increases (Fig. 4 lower left panel). However, increase in $k_{p300}$ forces $A_{p53}$ to reach amplitude death regime (apoptotic) quicker (Fig. 4 lower right panel). Similar scenario of transition of states is found in the case of $A_{Mdm2}$ (Fig. 5)\cite{ing}.

\subsection{Regulation of apoptosis}
Taming stress imparted in a system by stress induced parameters is important to save the system from apoptosis. The calculated critical value of $k_{SMAR1}$, $k^c_{SMAR1}$, at which the amplitude of $p53$ is zero, and larger than this value the system goes to apoptosis, corresponds to a value of $k_{HDAC1}$ for each $k^c_{SMAR1}$ (Fig. 6 upper panel). The phase diagram in the parameter space $(k_{HDAC1},k^c_{SMAR1})$ show the distinct demarcation of stress and apoptotic states (Fig. 6 upper panel). The result indicates that even for large value of $k_{SMAR1}$ which drives the sysytem to apoptotic state, one can vary $k_{HDAC1}$ so that the range of stress state be broaden such that the system can be pull back to normal state once the stress is removed. 

The average value of mid-value of sustain oscillation regime $p53$ amplitude ($A^{av}_{p53}$) for ten ensembles with different initial conditions modulated by $HDAC1$ as a function of $k_{SMAR1}$ shows monotonous decrease $A^{av}_{p53}$ as $k_{SMAR1}$ increases and will reach amplitude death for sufficiently large value of $k_{SMAR1}$ (Fig. 6 lower panel). Even though $k_{SMAR1}$ drives the $p53$ network to apoptosis (Fig. 6 lower panel), this apoptotic state can be regulated by $HDAC1$ interaction to save from apoptosis\cite{ale,naa}.

Similar study of the impact of $k_{SMAR1}$ on $p53$ network in regulating apoptotic phase in the presence of another stress inducer $p300$ via $k_{p300}$ shows different scenario. The apoptotic phase diagram in the parameter space $(k_{p300},k^c_{SMAR1})$ indicate two distinct scenarios, first $k^c_{SMAR1}$ increases as $k_{p300}$ increases upto a maxumim value, and secondly $k^c_{SMAR1}$ decreases as $k_{p300}$ increases (Fig. 7 upper panel). In the first case, for any critical value of $k^c_{SMAR1}$, one can extend the range of stress regime by increasing the concentration of $p300$ (increasing the value of $k_{p300}$) in the system and save the system from apoptosis after removing the stress. In the second case, the range of stress can be increased for any value of $k^c_{SMAR1}$ by decreasing concentration of $p300$ and can save from apoptotic state.

The value of $A^{av}_{p53}$ modulated by $p300$ decays slowly as a function of $k_{SMAR1}$ (exponential decay) as compared to the case of $HDAC1$ (Fig. 7 lower panel). The amplitude death scenario can be seen in this case also but with slow variation.
\begin{figure}
\label{fig7}
\begin{center}
\includegraphics[height=12.0cm,width=6.0cm]{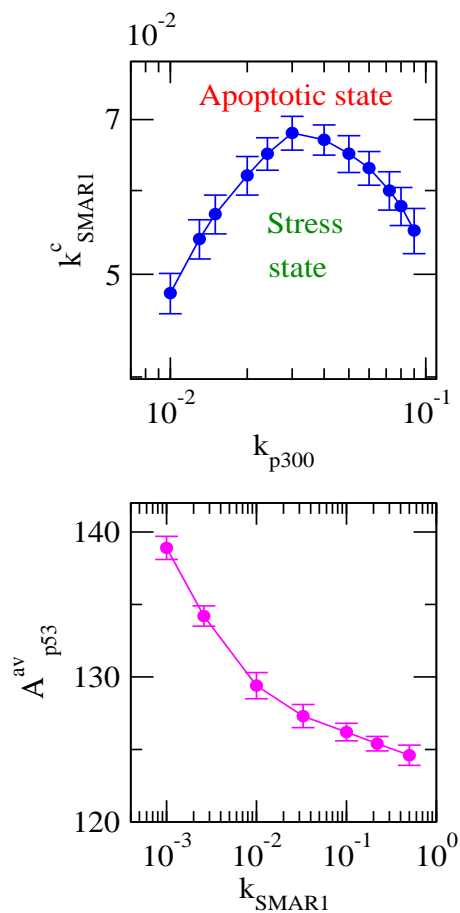}
\caption{(A) Phase diagram of $p53$ dynamics in the parameter space ($k_{p300},k^c_{SMAR1}$) showing distinct phases, namely, stress and apoptotic phases. (B) The plot of $A^{av}_{p53}$ as a function of $k_{SMAR1}$.} 
\end{center}
\end{figure}

\section{Conclusion}
Fluctuations imparted in a network, due to interaction of stress inducing molecular species in the form of a sub-network where the stress inducer species is the central in the sub-network, are propagated throughout the network and dynamical as well as topological properties of each individual component in the network get changed. However, the amount of perturbation signal in the form of stress recieved by various components in the network are not equal, and depend on how far the components are from the stress epicentre in the network. Biological network, corresponding to a certain biological function, is generally self-organized and tries to protect the network organization to maintain its own normal functioning. However, if the stress is large enough the network organization can no longer protect the network functioning from the stress and the functioning of the organization will break down and move to apoptosis.

$SMAR1$ is found to be a very dynamic and stress inducer signaling molecule which interfere the $p53$ regulatory network. It also interacts with many other signaling molecules such as $p300$, $HDAC1$ etc in the $p53$ network and regulate $p53$ dynamics. The concentration of this signaling molecular species in the network trigger the $p53$ dynamics to different states, which correspond to different cellular states, and even it can induce apoptosis to the cell. The mathematical modeling of this network provides various dynamical properties of the network which is reflected in the dynamics of the state vector which is the vector of molecular species variables in the system. The complexity of these states can be determined by calculating the permutation entropies of these states, and found that normal state corresponds to smallest value of permutation entropy. As stress increases, complexity also increases and permutation entropy is increased correspondingly, and surprisingly the permutation entropy of second stabilized state, which corresponds to apoptotic state, has highest value. This indicates that at apoptotic state the self-organization of the network has lost and become disorder in the network organization.

The amplitude death scenario obtained from the dynamical study of the $p53$ regulatory network model could be used as the signature of apoptosis. Because the dynamics of this state has large complexity due to the lost of self-organization at this state. On the other hand, the amplitude death for the case of normal state (stabilized state) has minimum complexity due to the maintainance of self-organization of the system. 

Abnormality in one signaling molecule in a system may trigger apoptosis to the system. However, since the network involves a number of other signaling molecules which can regulate the network, one can probably use other signaling molecules to save the system from apoptosis. The reason is that even though abnormality of one signaling molecule drives the system to apoptosis, a change in another signaling molecule may extend the range of stress and save the system from apoptosis. Thus even though $SMAR1$ can trigger apoptosis to $p53$ regulatory network, regulating other signaling molecule $p300$ or $HDAC1$ or both can possibly save the system from apoptosis. However, one needs experimental investigation and engineering of the signaling molecules in $p53$ regulatory network on such issues. Experimental and theoretical investigations in this direction are needed because these study will open up new understanding in the disease dynamics caused by abnormalities in signaling molecules, their preventive measures and cancer engineering.

\section*{Acknowledgments}
MZM is financially supported by Indian Council of Medical Research under SRF (Senior Research Fellowship). RKBS is financially supported by Department of Science and Technology (DST), New Delhi, India under sanction no. SB/S2/HEP-034/2012.


\end{document}